\documentstyle[aps,prl,twocolumn,epsf]{revtex}
\newcommand{\etal}{{\em et al.}}                

\newcommand{\eqref}[1]{(\ref{#1})}
\begin{document}
\title{
  \begin{flushright}{\rm DOE/ER/40561-53-INT}\\[9mm]\end{flushright}
  Imaging proton sources and space-momentum correlations
}
\author{Sergei Y. Panitkin$^{1}$ and David A. Brown$^{2}$ }
\address{\em $^1$Department of Physics, Kent
State University, Kent, Ohio 44242}
\address{\em $^2$Institute for Nuclear Theory, 
University of Washington, Box 351550, Seattle, WA 98195-1550}
\date{\today}
\maketitle
%
\begin{abstract}
The reliable extraction of information from the two-proton correlation 
functions measured in heavy-ion reactions is a long-standing problem.
Recently introduced imaging techniques give one the ability to reconstruct
source functions from the correlation data in a model independent way.  
We explore the applicability of two-proton imaging to realistic
sources with varying degrees of transverse space-momentum correlations.  
By fixing the freeze-out spatial distribution, we find 
that both the proton images and the two-particle 
correlation functions are very sensitive to these correlations.
We show that one can reliably reconstruct the source functions from
the two-proton correlation functions, regardless of the degree of the
space-momentum correlations. 
\end{abstract}
%
The sensitivity of the two-proton correlations to the space-time 
extent of nuclear reactions was first pointed out by
Koonin~\cite{koonin_77} and later emphasized by many
authors~\cite{nakai_81,lednicky_82,ernst_85,gong_91}.  
Since then, measurements of the two-proton 
correlations have been used along with pion HBT data as a probe of the 
space-time properties of the heavy-ion collisions (for the review of recent 
experimental results of two-particle interferometry
see~\cite{pratt_98,wiedemann_99,heinz_99} and references therein).  
A prominent ``dip+peak'' structure in the proton correlation
function is due to the interplay of the strong and Coulomb
interactions along with effects of quantum statistics. 
Because of the complex nature of the two-proton final state
interactions only model-dependent and/or qualitative statements were 
possible in proton correlation analysis.  Typically
~\cite{awes_95,aladin_fritz_97,aladin_schwarz_99,aladin_fritz_99,pp:NA49},
the proton source is assumed to be a chaotic source with gaussian
profile that emits protons instantaneously.  For simple static chaotic
sources, it has been shown~\cite{koonin_77,nakai_81,lednicky_82,ernst_85}
that the height of the correlation peak approximately scales inversely
with the source volume.  Heavy-ion collisions are complicated dynamic 
systems with strong space-momentum correlations (such as flow) and a 
nonzero lifetime.  Hence the validity of the assumptions behind such 
simplistic sources is questionable.  In order to address the limitations 
of this type of analysis (and to incorporate collective effects), some 
authors~\cite{gong_91,pp:NA49,pp:e877} utilize transport models
to interpret the proton correlation functions.  Although this
approach is a step in a right direction, it is still highly model-dependent.\\
\indent Recently, it was shown that one can perform model-independent
extractions  of the {\em entire} source function 
$S(r)$ (the probability density for emitting protons a distance $r$ apart)
from two-particle correlations, not just its
radii, using imaging techniques
\cite{HBT:bro97,HBT:bro98,HBT:bro98a,Danielewicz:1998kp}. 
Furthermore, one can do this even with the relatively complicated
proton final-state interactions and without making any {\em a priori} assumptions
about the source geometry or lifetime, etc.
First results from the application of imaging to the
proton correlation data can be found in refs.~\cite{HBT:bro97,HBT:bro98,pp:e895}. 
While these results look promising, tests of the imaging technique have  
only been performed on static (Gaussian and non-Gaussian) sources.
It is important to understand the limitations and robustness of this 
technique especially in the light of the ongoing experimental program
at SIS, AGS and SPS as well as upcoming experiments at RHIC.\\
\indent In this letter, we will study the applicability of proton
imaging to realistic sources with transverse space-momentum
correlations.
In particular we explore how $\vec{r}_T-\vec{p}_T$ correlations
{\em directly} affect the proton sources and,
hence, the shapes of the experimentally observable correlation functions. 
Here $\vec{r}_T$ and $\vec{p}_T$ are the transverse radius and transverse
momentum vectors respectively of a proton at the time when it decouples
from the system (freeze-out).
It has been argued in the pion~\cite{pratt84} and
proton~\cite{gong_91} cases that the apparent source size (or the effective 
volume) decreases as collective motion increases.   
We will verify this expectation and show that one can reliably 
reconstruct the source function, even in the presence of extreme
space-momentum correlations.
The outline of this letter is as follows.  First, we briefly describe
the imaging procedure used to extract the source function from experimental 
correlations. We will discuss proton
sources but most of our arguments and conclusions are valid for
any two-particle correlations. Next, we describe how we implement the
varying degrees of space-momentum correlations using the RQMD model.
 Finally, we will discuss the influence of these correlations on the
proton correlation functions and imaged sources.  Since we can also
construct the sources directly within RQMD, this serves as a more
demanding test of the imaging procedure than has been performed to
date.\\
%
\indent With imaging, one extracts the entire source function $S(r)$ from the 
two-proton correlation function, $C(q)$.  Here the source function is
the probability density for emitting protons with a certain relative
separation {\em in the pair  Center of Mass (CM) frame}.  The source function
and the correlation 
function are related by the equation \cite{koonin_77,HBT:pra90}:
\begin{equation}
  C(q)-1=4 \pi \int_0^\infty dr r^2 K(q,r) S(r)
  \label{eqn:pk}
\end{equation}
In eq.~\eqref{eqn:pk}, $q=\frac{1}{2}\sqrt{(p_1-p_2)^2}$ is the 
invariant relative momentum of the pair, $r$ is the pair CM separation 
after the point of last collision, and $K$ is the kernel.  The kernel 
is related to the two-proton relative wavefunction via
\begin{equation}
  K(q,r) = \frac{1}{2} \sum_{js\ell\ell'}(2j+1) 
  \left(g_{js}^{\ell\ell'}(r)\right)^2-1
  \label{eqn:kern}
\end{equation}
Here $g_{js}^{\ell\ell'}$ are the relative proton radial wavefunctions for 
orbital angular momenta~$\ell, \ell'$, total angular momentum~$j$, and total 
spin~$s$.  In what follows, we calculate the proton
relative wavefunctions by solving the Schr\"odinger equation with the 
REID93~\cite{sto94} nucleon-nucleon and Coulomb potentials.

Because~\eqref{eqn:pk} is an integral equation with a non-singular kernel, 
it can be inverted~\cite{HBT:bro97,HBT:bro98}.  To perform the 
inversion, we first discretize eq.~\eqref{eqn:pk}, giving a set of linear
equations, $C_i-1=\sum_{j=1}^MK_{ij}S_j$, with $N$ data points and $M$ source 
points.  Given that the data has experimental error $\Delta C_i$, one cannot 
simply invert this matrix equation.  Instead, we search for the source vector 
that gives the minimum $\chi^2$:
\begin{equation}
  \chi^2=\sum_{i=1}^{N} \frac{(C_i-1-\sum_{j=1}^MK_{ij}S_j)^2}
  {(\Delta C_i)^2}.
  \label{eqn:chi2}
\end{equation}
The source that minimizes this $\chi^2$ is (in matrix notation):
\begin{equation}
  S=(K^{\rm T}BK)^{-1}K^{\rm T}B(C-1)
  \label{eqn:answer}
\end{equation}
where $K^{\rm T}$ is the transpose of the kernel matrix and 
$B$ is the inverse covariance matrix of the data,
$B_{ij}=\delta_{ij}/(\Delta C_i)^2$.
In general, inverse problems such as this one are {\em ill-posed}
problems.  In practical terms, small fluctuations in the data can lead 
to large fluctuations in the imaged source.  
One can avoid this problem by using the method of 
Optimized Discretization discussed in reference \cite{HBT:bro98}.
In short, the Optimized Discretization method varies the size of the 
$r$-bins of the source (or equivalently the resolution of the kernel) 
to minimize the relative error of the source.  

The source function that one reconstructs is directly related to the 
space-time development of the heavy-ion reaction in the Koonin-Pratt
formalism \cite{koonin_77,HBT:pra90,neq:dan92}:
\begin{equation}\begin{array}{rl}
  S(r,\vec{q})=& \displaystyle  
  \int_{4\pi} d\Omega_r \int dt_1 dt_2 \int d^3R \\ 
  &\displaystyle \times D(\vec{R}+\vec{r}/2,t_1;\vec{q})
  D(\vec{R}-\vec{r}/2,t_2;-\vec{q}),
\end{array}\label{eqn:defofsource}\end{equation}
where $\vec{q}=\frac{1}{2}(\vec{p}_1-\vec{p}_2)$, making $q=|\vec{q}|$.  
Here the $D$'s are the normalized single particle sources in the pair CM 
frame and they have the conventional interpretation as the normalized phase-space 
distribution of protons after the last collision (freeze-out) in a transport 
model.  In computing $S(r,\vec{q})$ in a transport model, 
one does not need to consider the contribution of large relative momentum 
($q\gtrsim q_{\rm cut}$) pairs to the source as the kernel cuts off the
contribution from these pairs.  The kernel does this because it 
is highly oscillatory while the source varies weakly on the scale of these 
oscillations and the integral in \eqref{eqn:pk} averages to zero. We
can estimate $q_{\rm cut}$ directly from the correlation function  
as $q_{\rm cut}$ is roughly the momentum where the correlation goes to
one.  Nevertheless, for the imaging in~\eqref{eqn:pk} to be unique 
one must require that the $q$ dependence of the correlation comes from the 
kernel value alone and eq.~\eqref{eqn:defofsource} seems to indicate that the 
source itself
has a $q$ dependence.  In practice $S(r,\vec{q})$ only has a weak $\vec{q}$ 
dependence for $q\lesssim q_{\rm cut}$ and this dependence may be 
neglected~\cite{HBT:pra90,gong_91}.\\
\indent Since $S(r)$ is the probability density for finding a pair 
with a separation of emission points $r$, one can compute it
directly from the freeze-out phase-space distribution given by some model.   
First one scans through this freeze-out density of protons, 
then histograms the number of pairs in relative distance in the CM, and finally 
normalizes the distribution: $4\pi \int dr r^2 S(r)=1$.
As mentioned above, only low relative momentum pairs may enter into this 
histogram as the kernel cuts off the contribution from pairs with 
$q > q_{\rm cut}$. \\
%
\indent In our studies we used the Relativistic Quantum Molecular
Dynamics (RQMD) model~\cite{sorge_95}.
It is a semi-classical microscopic model which includes stochastic
scattering, classical propagation of the particles. It includes
baryon and meson resonances, color strings and some quantum effects
such as Pauli blocking and finite particle formation time. This model
has been successfully used to describe  many features of relativistic
heavy-ion collisions at AGS and SPS energies. 
Our approach is as follows: first we take the
freeze-out phase space distributions generated by RQMD and we alter
the orientation of transverse momentum relative to the transverse
radius, obtaining a subset of the phase space points.
Following this, we use the Lednicky-Lyuboshitz~\cite{lednicky_82,gmitro86} method 
to construct the proton-proton correlation function. This method gives a
description of the final state interactions between two protons, including
antisymmetrization of their relative wave function. Finally, using the
imaging technique described above we compute the
proton source functions. We used $\approx 4000$ simulated events
of $4$~GeV/A Au-Au reactions with impact parameter $b\leq 3$~fm.  We
utilized only pairs in the central rapidity region with $|y| \leq
0.3$ and applied no cut on transverse momentum $p_T$.\\
\indent We consider three different degrees of alignment between the transverse
position $\vec{r}_T$ and the transverse momentum $\vec{p}_T$ of each proton 
used to construct the correlation function. These alignments are 
implemented in the same manner as in reference \cite{monreal99}:
\begin{enumerate}
\item We orient $\vec{p}_T$ at a random angle with respect to $\vec{r}_T$. 
We refer to this as the random case.  One can think of this case
as being ``thermal'' as the transverse flow component is completely
removed.
\item We do not change the orientation of $\vec{p}_T$.  We refer to this case as the 
unmodified case.
\item We align $\vec{p}_T$ with $\vec{r}_T$ and refer to this as
the aligned case.  One can think of this case as one of extreme transverse 
flow.
\end{enumerate}
Note that the rotation occurs in the rest frame of the colliding
nuclei.  In all cases, we only rotate $\vec{p}_T$ so these procedures do not change 
the spatial distribution at freeze-out.  However, it is clear that these procedures do
change the phase-space density.\\  
%
    \begin{figure}

        \epsfxsize=3.4in 
        \epsffile{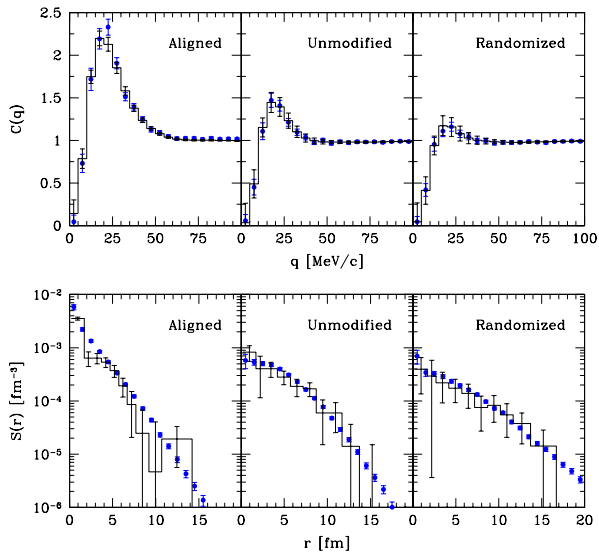}
      \caption[]{Upper panels: Original (solid symbols) and restored
      (histogram)  
      proton correlation functions for different degrees of
      space-momentum alignment. The error bars in the original
      correlation functions are from the limited 
      statistics of the RQMD runs.  Lower panels: Reconstructed
      sources (histogram) and source computed directly in RQMD (solid symbols)
      for the different degrees of alignment.} 
      \label{fig:result}
    \end{figure}
Upper panels on Fig.~\ref{fig:result} show the correlation 
functions for the three different cases.
It is clear from the Figure that the degree of space-momentum
correlation has a strong  
influence on the correlation function: the peak height of the
correlation function changes from about 1.45 for unmodified RQMD to
about 2.2 for the aligned case and 1.2 for randomized case.    
We would like to stress again, in all cases the spatial part of the source 
e.g. ``radius'' or ``volume,'' remains unaltered
as does the transverse momentum spectrum and rapidity
distribution of protons.  Hence, the upper panels of Fig. 1 illustrate the danger
of ignoring space-momentum correlations when analyzing correlation data.\\
\indent On lower panels in Fig.~\ref{fig:result} we show the
proton sources obtained with the help of the imaging procedure outlined
above. Notice that, as the degree of alignment increases
(going from right to left), that the source function becomes
narrower and higher. \\
    \begin{figure}
      \begin{center}
        \epsfxsize=3.5in 
        \epsffile{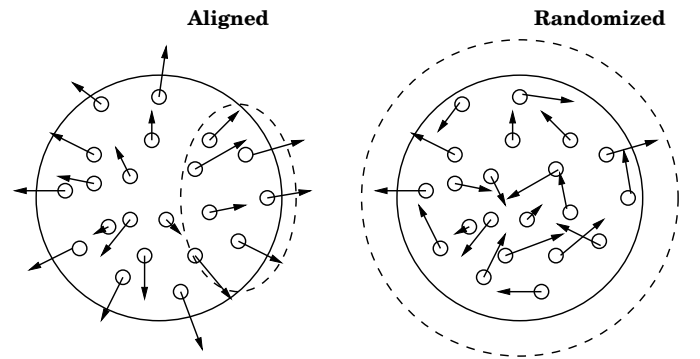}
      \end{center}
      \caption[]{Sketch of single particle sources at central rapidity, 
looking in the beam direction for the aligned case (left) and random case
(right). The small circles represent the protons and the arrows
represent their transverse momentum vectors.  The areas outlined with
the dashed lines represent regions where we find pairs with small
relative momentum.} 
      \label{fig:sketch}
    \end{figure}
One can understand this shift to lower separations in the way sketched in 
Fig.~\ref{fig:sketch}.  In the 
aligned case, it is more probable that nearby protons have a small
relative momentum $q$.  In the 
random case, any pair can have a small $q$, regardless of their separation.  
Given that the kernel cuts off contributions from pairs with larger $q$, we 
expect that the aligned case will have a narrower source than the unmodified 
case and the unmodified case will have a narrower source than the random case.
Also, in Fig.~\ref{fig:result} we have shown the sources constructed 
directly from the RQMD freeze-out distribution following
eq.~(\ref{eqn:defofsource}). In these sources, we considered all pairs with a 
relative momentum smaller than $q_{\rm cut}= 60$ MeV/c.  We explored
a range of  $q_{\rm cut}$ of 60 to 100 MeV/c, all beyond the
point in the  correlation where it is consistent with one, and found no cutoff
dependence.\\ 
\indent In all cases we see a general agreement of the imaged sources with the low
relative momentum sources constructed directly from RQMD.  In order to 
check the quality of the imaging and numerical
stability of the inversion procedure, the two-proton correlation
functions were calculated using the extracted relative source functions
shown on lower panels in Fig.~\ref{fig:result} as an input for
eq.~\eqref{eqn:pk}. The result of such ``double 
inversion'' procedure is shown on upper panels in
Fig.~\ref{fig:result} with solid circles. The agreement between the
measured and reconstructed correlation function is quite good,
confirming that imaging produces numerically stable and unbiased
results.\\  
\indent In conclusion, we have explored the applicability of proton
imaging to realistic sources with transverse space-momentum correlations.  
By fixing freeze-out spatial distribution and varying the 
degree of transverse space-momentum correlation we found that both 
the images and the two-particle correlation functions are very sensitive 
to these correlations.
In particular, we have shown that the source function narrows
(i.e. the probability of emitting pairs with small relative
separation grows) and the peak of the
proton correlation function increases as the degree of alignment
increases. 
Finally, we have demonstrated that one can reliably reconstruct the source
functions even with extreme transverse space-momentum correlations.
We would like to point out that the effects of space-momentum
correlations should be even more pronounced 
in the shapes of three-dimensional proton sources.   
Note that three dimensional proton imaging is now possible \cite{APSmeeting}.\\
\indent An important direction for the future is a detailed study of
the change of the phase-space density and entropy (extracted from
imaged sources~\cite{HBT:bro97}) with the varying degree of
space-momentum correlation. 
Such work should provide information complementary to ongoing studies
in the pion sector \cite{Ferenc:1999ku}.\\
%
\indent We gratefully acknowledge stimulating discussions with 
Drs. G.~Bertsch, P.~Danielewicz, D.~Keane, A.~Parre{\~n}o, S~Pratt,
S.~Voloshin and N.~Xu. We also wish to thank Drs.~R.~Lednicky and
J.~Pluta for making their correlation afterburner code
available. Finally, we thank Dr. H.~Sorge for providing the code of
the RQMD model.  This research is supported by the U.S. Department of
Energy grants  DOE-ER-40561 and DE-FG02-89ER40531. 
%

\end{document}